\documentclass[runningheads]{llncs}
\usepackage[T1]{fontenc}
\usepackage{graphicx}
\usepackage{booktabs}
\begin{document}
\title{Will You Participate? Exploring the Potential of Robotics Competitions on Human-centric Topics}
%
%
\author{Yuchong Zhang\inst{1}\orcidID{0000-0003-1804-6296} \and
Miguel Vasco\inst{1}\orcidID{0000-0002-5761-4105} \and
M\aa{}rten Bj{\"o}rkman\inst{1}\orcidID{0000-0003-0579-3372} \and
Danica Kragic\inst{1}\orcidID{0000-0003-2965-2953}}
\authorrunning{Y. Zhang et al.}
%
\institute{KTH Royal Institute of Technology, Stockholm, Sweden
\email{\{yuchongz,miguelsv,celle,dani\}@kth.se}}
\maketitle              
\begin{abstract}
This paper presents findings from an exploratory needfinding study investigating the research current status and potential participation of the competitions on the robotics community towards four human-centric topics: \textit{safety}, \textit{privacy}, \textit{explainability}, and \textit{federated learning}. We conducted a survey with 34 participants across three distinguished European robotics consortia, nearly 60\% of whom possessed over five years of research experience in robotics. Our qualitative and quantitative analysis revealed that current mainstream robotic researchers prioritize \textit{safety} and \textit{explainability}, expressing a greater willingness to invest in further research in these areas. Conversely, our results indicate that \textit{privacy} and \textit{federated learning} garner less attention and are perceived to have lower potential. Additionally, the study suggests a lack of enthusiasm within the robotics community for participating in competitions related to these topics. Based on these findings, we recommend targeting other communities, such as the machine learning community, for future competitions related to these four human-centric topics.

\end{abstract}

\section{Introduction}
\label{intro}

\begin{figure}[ht!]
  \includegraphics[width=\columnwidth]{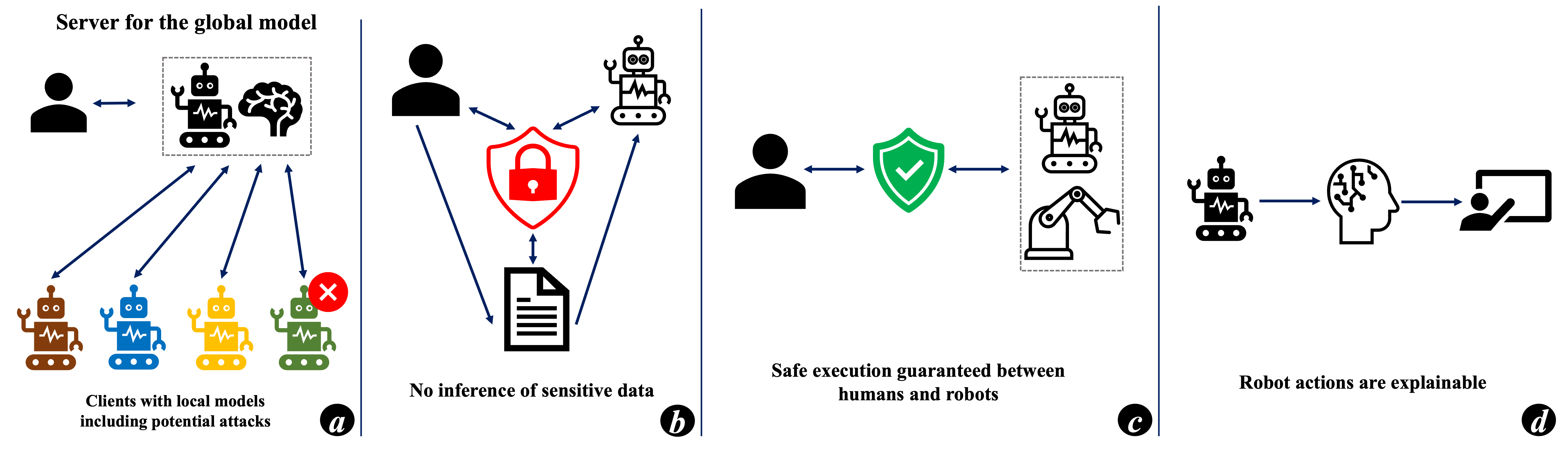}
  \caption{The four human-centric topics identified in this paper. $a:$ Federated learning; $b:$ Privacy; $c:$ Safety; $d:$ Explainability.}
  \label{fig:teaser}
\end{figure}

Robotics-themed competitions have been successfully initiated and consistently held spanning several decades, along with numerous prestigious conference venues like ICRA, IROS, NeurIPS, HRI, etc \cite{dias2016robot}. The topics of the competitions are diverse across a wide array of application domains. In recent years, the focus of robotics competitions has gradually evolved beyond purely pragmatic techniques to incorporate human-centric perspectives \cite{parreira2023did}, including some aspects of human-robot proxemics \cite{samarakoon2022review}. This paper aims to gain a comprehensive understanding of the potential impact of four specific and widespread human-centered topics -- \textit{safety}, \textit{privacy}, \textit{explainability}, and \textit{federated learning} -- on robotics competitions as shown in Figure ~\ref{fig:teaser}. These topics were determined in line with a Europe-wide project -- (ELSA) European Lighthouse on Secure and Safe AI (\url{https://benchmarks.elsa-ai.eu/?ch=5\&com=introduction}). We conducted the first (to our knowledge) exploratory study to assess the current status, underlying interests, and future prospects of these topics within the perspectives of robotics/human-robot interaction (HRI) communities. Through a needfinding study \cite{chung2018your,zhang2023industrial,chung2018exploring}, we sought to gather the feedback and insights and from robotics researchers and practitioners regarding these four specialized topics, so as to offer inspiration and guidance for future competitions.

\emph{Safety} is a fundamental consideration when designing robotic systems, in particular for interaction scenarios with humans. These systems haev the potential to inflict physical harm on humans, due to the forces exerted on humans during direct contact with the robot, and psychological harm, due to repeated violation of social norms and conventions during interaction with the robot~\cite{lasota2017survey}. This risk has lead to significant efforts to develop international safety standards, specifically for human-robot collaboration, which define quantitative biomechanical limits on the forces exerted on different parts of the human body and monitoring tools for robot motion~\cite{iso2016robots}. In literature, safety can be achieved through different methods: through control, developing methods that limit the velocity and motion of the robot prior to and post collision with the human user; through motion planning, developing methods to plan safe robot paths and motions that avoid potential collisions with humans; though prediction, developing methods that either predict human actions and motions to anticipate potential collisions; and through considering psychological factors, developing methods to monitor the quality of the interaction and identifying which factors influence the perceived safety of the robot~\cite{lasota2017survey}.

\emph{Privacy} is another emerging topic in HRI research, due to their increasing appearance in multiple social contexts with humans. In particular, the majority of literature has focused on three particular privacy phenomena of the interaction with social robots~\cite{lutz2019privacy}. The first concerns the surveillance capabilities of social robots, due to their potential autonomous mobility, that can affect the physical privacy of human users (e.g., entering autonomously a private room). The second concerns the social bonding capabilities of social robots, that may lead the human users to disclose private information. Finally, the third phenomena concerns the opacity of the processing and data collection capabilities of social robots, which may be misunderstood by the human user.

\emph{Explainability} is a traversal topic to AI that attempts to build systems that are able to find human-interpretable interpretations for complex pattern recognition models. In the context of autonomous social robots, the goal is build systems that are able to explain its actions to human users while interacting in a shared space~\cite{sakai2022explainable}. To achieve this goal, four steps have been proposed~\cite{sakai2022explainable}: designing or learning a robot decision-making space that is interpretable by a human; estimating the human decision-making space and its planning algorithm, which allows the generation of personalized explanations; extracting information that is important for communicating plans; converting explanatory factors into an efficient medium (either speech of visual information).

\emph{Federated learning} (also referred to as collaborative learning), unlike the conventional machine learning, is an advanced machine learning methodology that involves training an algorithm through numerous independent sessions, each utilizing its distinct dataset \cite{he2023machine}. Having garnered significant attention in recent years, this methodology diverges from traditional centralized machine learning approaches where local datasets are amalgamated into a single training session, and it differs from methods assuming identical distribution of local data samples. Federated learning empowers multiple participants to collaboratively develop a shared and resilient machine learning model without the need to disclose individual datasets. This approach effectively tackles crucial concerns, including data privacy, data security, and data access rights \cite{li2020review}. Yet, data privacy is the most prevailing topic associated with federated learning since it allows building personalized models without violating sensitive information, complying with ethical principles \cite{zhang2021survey,rieke2020future,xu2021federated}. Aligning with the common acknowledgement of human-computer interaction (HCI): It aims to design a seamless connection between users and machines for required services, optimizing performance in terms of quality and efficiency \cite{karray2008human}. Therefore, federated learning exhibits a notable connection with HCI where this intersection becomes imperative in line with the human-centric design considerations and ethical implications in privacy-preserving machine learning systems \cite{chhikara2020federated}. Specifically, federated learning has already been employed for improved distributed learning in HRI scenarios \cite{gamboa2023asynchronous}.

\begin{figure}[ht!]
    \centering
    \includegraphics[width=.7\columnwidth]{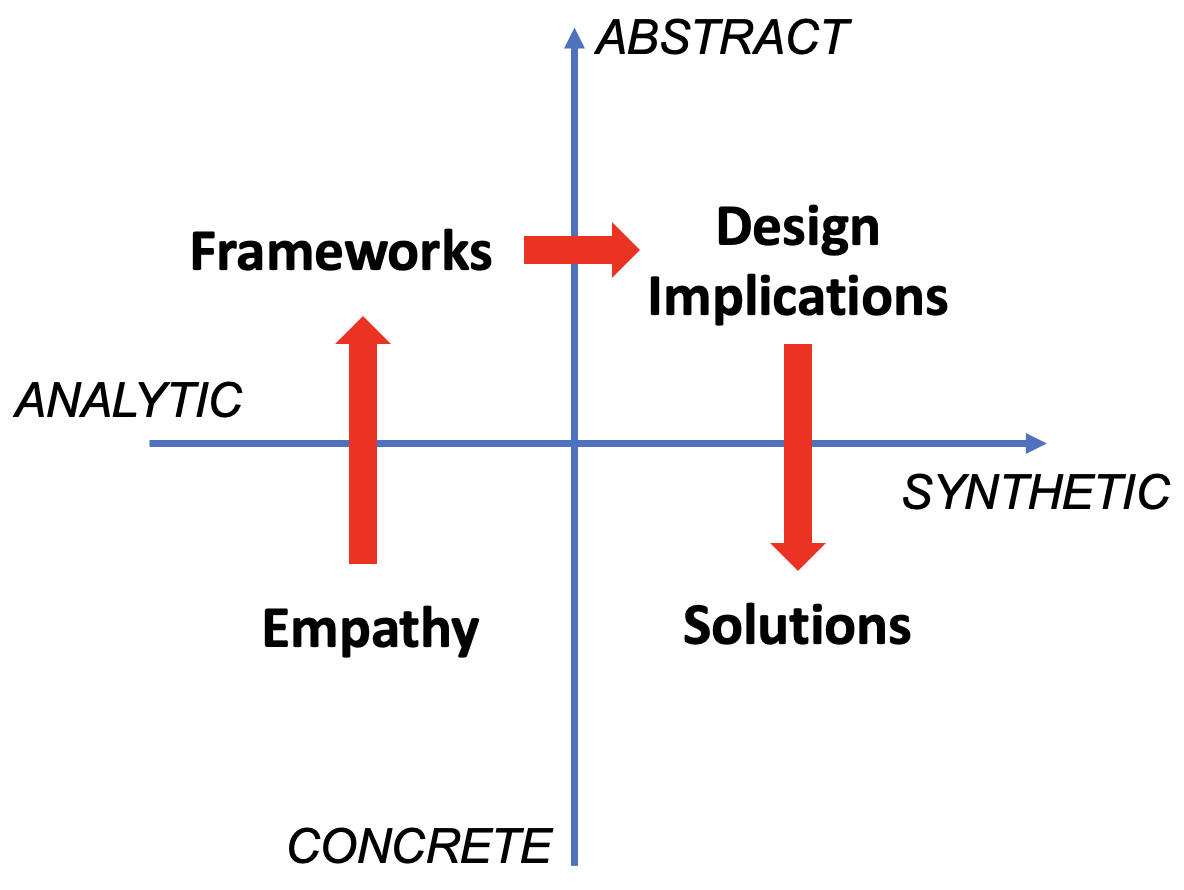}
    \caption{The general overview of a needfinding process.}
    \label{fig:enter-label}
\end{figure}

We employed a method known as 'needfinding', derived originally from the research process product design \cite{beckman2007innovation}. The objective of this methodology is to pinpoint the essential user needs that a product is intended to fulfill towards the specific group. The general needfinding process is outlined in Figure ~\ref{fig:enter-label} \cite{pantofaru2012exploring}. It commences by fostering empathy for the user group (via interviews, surveys, or other media) and subsequently the outcomes from this steps will be extracted and presented in frameworks, for instance, graphical representations. Then, the design implications are formulated from the framework while the ultimate solutions are fabricated. In our paper, the empathy stands for designing a goal-oriented survey targeted for the robotics community, while the frameworks represent the extracted results from the survey. Specifically, the approaches we used for obtaining the outcomes combined both qualitative and quantitative approaches. In the qualitative phase, we summarized and coded the responses we've gathered to assess the prevailing circumstances, future aspirations, and the envisioned benefits reported by participants. This qualitative analytics provided valuable insights into the perspectives to be used by the robotics/HRI community towards the four topics in future robotics competitions. In the quantitative analysis, we conducted statistical significance tests on select survey variables that are expressed on mathematical scales. By employing these statistical methods, we aimed to uncover patterns, trends, and the importance of different factors related, thus providing a second viewpoint which is objective data-driven understanding of the surveyed results. The design implications are revealed as findings with insights formulated from the results, employed as guidelines (solutions) for future considerations in robotics competitions related to these human-centric topics. In our paper, we affirmed the lack of a strong inclination within the robotics community to participate in these competitions. We also suggest the potential for a shift of the target group, such as the machine learning community.

The organization of this paper is as follows. Section ~\ref{intro} introduces the basic background and emphasizes the principles of needfinding of out study. Section ~\ref{rw} presents the related work while Section ~\ref{meth} delineates the employed methodology, including the design strategy of our questionnaire. The results, comprising both quantitative and qualitative analyses are displayed in Section ~\ref{res}, followed by an in-depth discussion presented in Section ~\ref{dis}. Section ~\ref{con} concludes the paper.  

\section{Related Work}
\label{rw}
A literature review was conducted, revealing that needfinding (usually in the form of surveys, interviews, workshops, etc) has been extensively used in exploring the needs of target users within a broad field of robotics and HRI research \cite{pantofaru2011need,zubrycki2016understanding,martelaro2017dj}. Chung et al. \cite{chung2018exploring} designed four case studies including survey and interview to probe the potential usage of physical robots in collecting feedback in hotels. Especially, they were aware of the privacy of guests during the interaction with the service robots. O'Brien et al. \cite{o2021exploring} employed a comprehensive needfinding process in prototyping a therapeutic robot for children, where involved a detailed literature crawling, an ethical analysis, and interviews and pediatric experts. Similarly in designing a robotic system for children, Bejarano et al. \cite{bejarano2021designing} executed a semi-structured needfinding interview with local experts in children's hospital to identify the requirements of robots to be empathetic and emotionally supportive. Vatsal et al. \cite{vatsal2017wearing} engaged a needfinding contextual inquiry to generate guidance before designing specific capabilities for a wearable robotic forearm. Apparently, the utilization of needfinding has been proven its efficacy in gathering early-stage feedback and requirements before delving into the practical artifacts. However, no prior work, to our knowledge, has investigated the correlation of human-centric topics with robotic competitions, which formed the motivation of our paper.

\section{Research Question}
Our study aims to assess the present research landscape and the potential of competition participation aligned with the four human-centric topics on the robotics community. Our objective is to disclose the current and prospective interests of the engagement of robotics competitions so as to formulate the guidelines of hosting future events for researchers and practitioners. The research questions addressed in this paper are listed below:

\begin{itemize}
    \item What are the current interests situated in robotics research on \textit{safety}, \textit{privacy}, \textit{explainability}, and \textit{federated learning} in terms of the robotics/HRI community?
    \item What are the willingness of the robotics/HRI community in participating robotics competitions related to \textit{safety}, \textit{privacy}, \textit{explainability}, and \textit{federated learning} and the potential benefits or drawbacks brought afterwards?
\end{itemize}

\begin{figure*}[ht]
  \centering
  \includegraphics[width=\columnwidth]{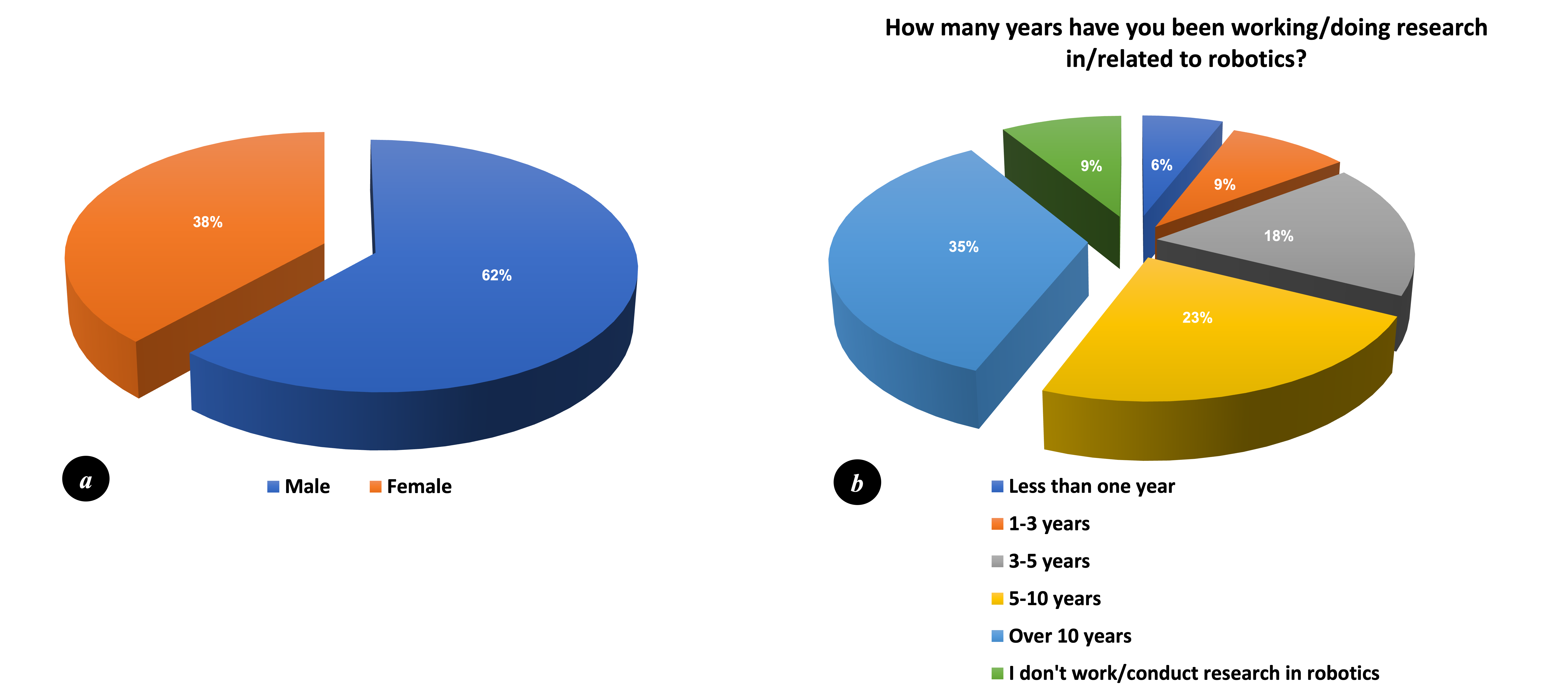}
  \caption{Background information of participants. $a:$ Gender information. $b:$ Academic background of participants.}
  \label{fig:parti}
\end{figure*}

To address these questions, we conducted a survey study in the form of a crafted questionnaire containing both closed and open-ended questions targeted for the robotics community. By concentrating on this distinct group, our aim is to derive valuable insights and draw conclusions regarding the intent and willingness of robotics researchers/practitioners to participate in competitions centered around the four topics.

\section{Method}
\label{meth}
In this section, we introduce the approach used in our paper, including the participants recruited and the specification of the survey designed.

\subsection{Participants}
We conducted our survey study with a questionnaire including a number of curated questions which were oriented exclusively for the robotics community. We invited 34 participants (21 self-identified males and 13 self-identified females) by emailing from three distinct and reputable robotics consortia: (1) a distinguished Europe-wide project community euRobin (\url{https://www.eurobin-project.eu/}) specializing in advanced and reproducible research in robotics and (2) the local academic division (including faculty members, research fellows, PhD students, research engineers) of robotics of the authors' university and (3) the renowned robotics company PAL Robotics (\url{https://pal-robotics.com/}) to complete the survey. Their ages ranged from 23 to 54 (mean = 35.68, SD = 8.43). As shown in Figure ~\ref{fig:parti}.$b$, nearly 60\% of participants possess hands-on experience in research or work within the field of robotics, spanning over a noteworthy period of more than 5 years. Furthermore, the highest proportion (35.3\%) is observed among those with an extensive background, surpassing 10 years of experience in robotics. All participants approved a consent that all of the collected data (no private and sensitive information) from the survey will be kept confidential and merely used for research.

\begin{figure}[ht!]
  \centering
  \includegraphics[width=\columnwidth]{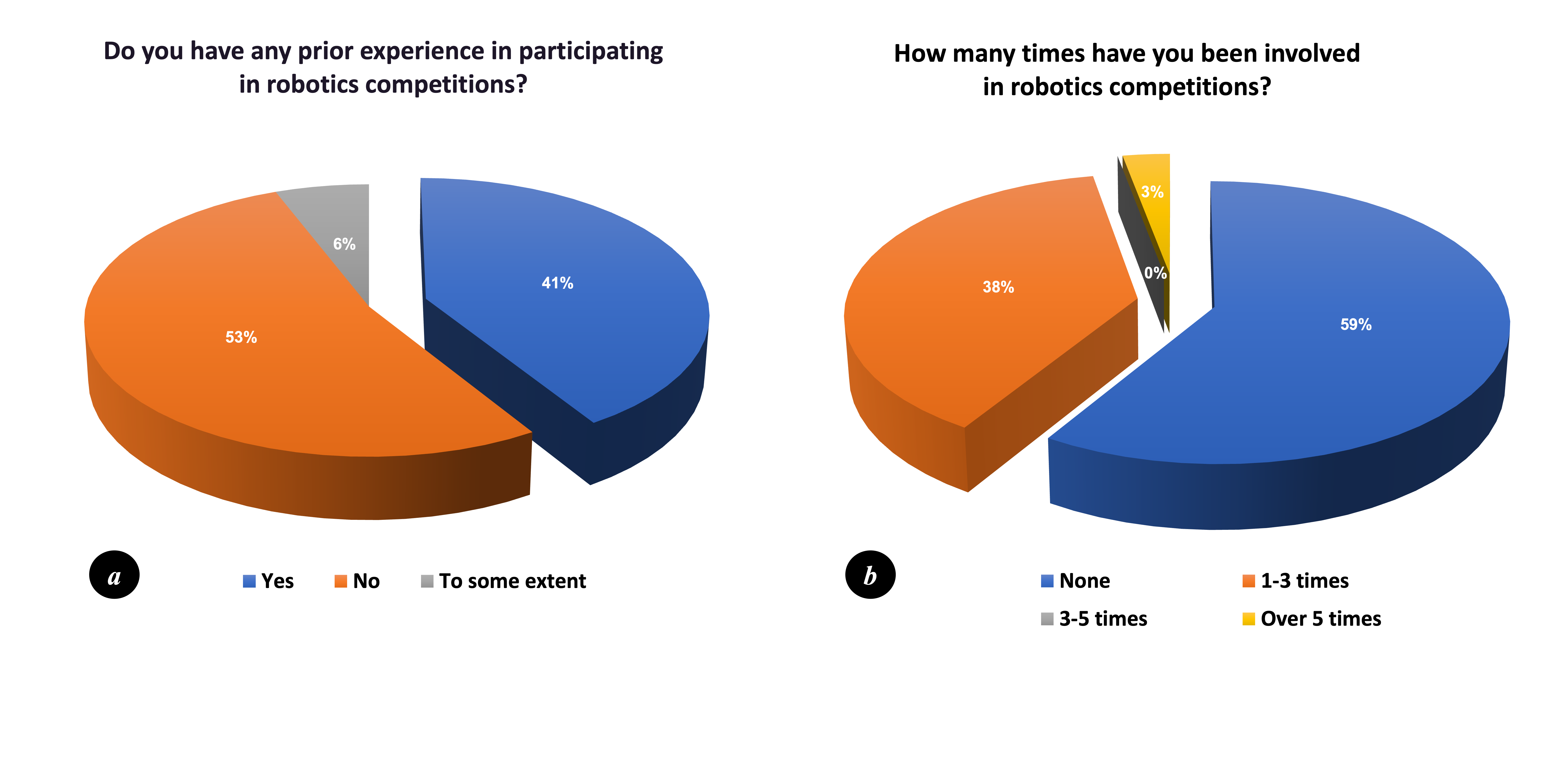}
  \caption{Prior information of participants' past involvement in robotics competitions.}
  \label{fig:prior}
\end{figure}

\subsection{Questionnaire}
The questionnaire was rigorously crafted through numerous iterative discussions among the authors, ensuring its representative nature. First, we extracted basic demographic information, including gender, age, and years of experience. Then we investigated information about the prior experience of robotics completions. A shown in Figure ~\ref{fig:prior}.$a$, over half of participants have zero experience in robotics competitions (52.9\%). Specifically, 5.9\% of participants reported that they have been involved in robotics competitions "to some extent", from that we believe this signifies informal involvement in certain aspects of the competition procedure, rather than any deliberate, structured participation. Interesting, this complied with the information that 41.2\% of participants indicated preceding involvement with robotics competitions on at least one occasion (Figure ~\ref{fig:prior}.$b$). Next, we concentrated on identifying the current research landscape, the future possibilities, and the perceived importance of the four topics in robotics and competitions as far as the community concerned. As depicted in Table ~\ref{fig:likert}, we presented a set of questions to gauge the extent and intent of exploration within the robotics community regarding the four tested topics. These queries were formulated using a 7-point Likert scale to quantitatively assess the perceived responses. Finally, we provided a few objective and open-ended questions, aiming to gather the opinions of the participants towards these four topics. Some extra short follow-up interview sessions were complemented as needed. Finally, we verified that all responses from the 34 participants were deemed valid.

\begin{table}[ht!]
\centering
  \caption{The questions (\textit{\textbf{Q1--Q5}}) quantified by 7-point Likert sale.}
  \label{ques}
  \begin{tabular}{cll}
    \toprule
      & Questions & Scale: 1--7 \\
    \midrule
    \textit{\textbf{Q1}} & To what extent do you explore Federated & Very inactive--Very active \\
     & Learning, Privacy, Safety, and Explainability & \\
     & in your research? & \\
    \textit{\textbf{Q2}} & Do members of your research group & Very unlikely--Very likely \\
     & or lab or division explore the topics of & \\
     & Federated Learning, Privacy, Safety, & \\
     & and Explainability in Robotics? & \\
    \textit{\textbf{Q3}}& Are Federated Learning, Privacy, Safety, and & Very unlikely--Very likely \\
     & Explainability in robotics potential research & \\
     & fields that you would like to explore in your & \\
     & future research? & \\
    \textit{\textbf{Q4}}& Would you have interest in participating & Very unlikely--Very likely \\
     & in a robotics competition focused on the & \\
     & topics of Federated Learning, Privacy, & \\
     & Safety, and Explainability? & \\
    \textit{\textbf{Q5}}& How do you rate the importance of Federated & Very unimportant--Very important \\
     & Learning, Privacy, Safety, and Explainability & \\
     & in robotics?  & \\
    \bottomrule
  \end{tabular}
\end{table}

\section{Results}
\label{res}
In this section, we present the results obtained, including the quantitative and qualitative outcomes.

\subsection{Quantitative Analysis}
\begin{figure}[ht!]
  \centering
  \includegraphics[width=\columnwidth]{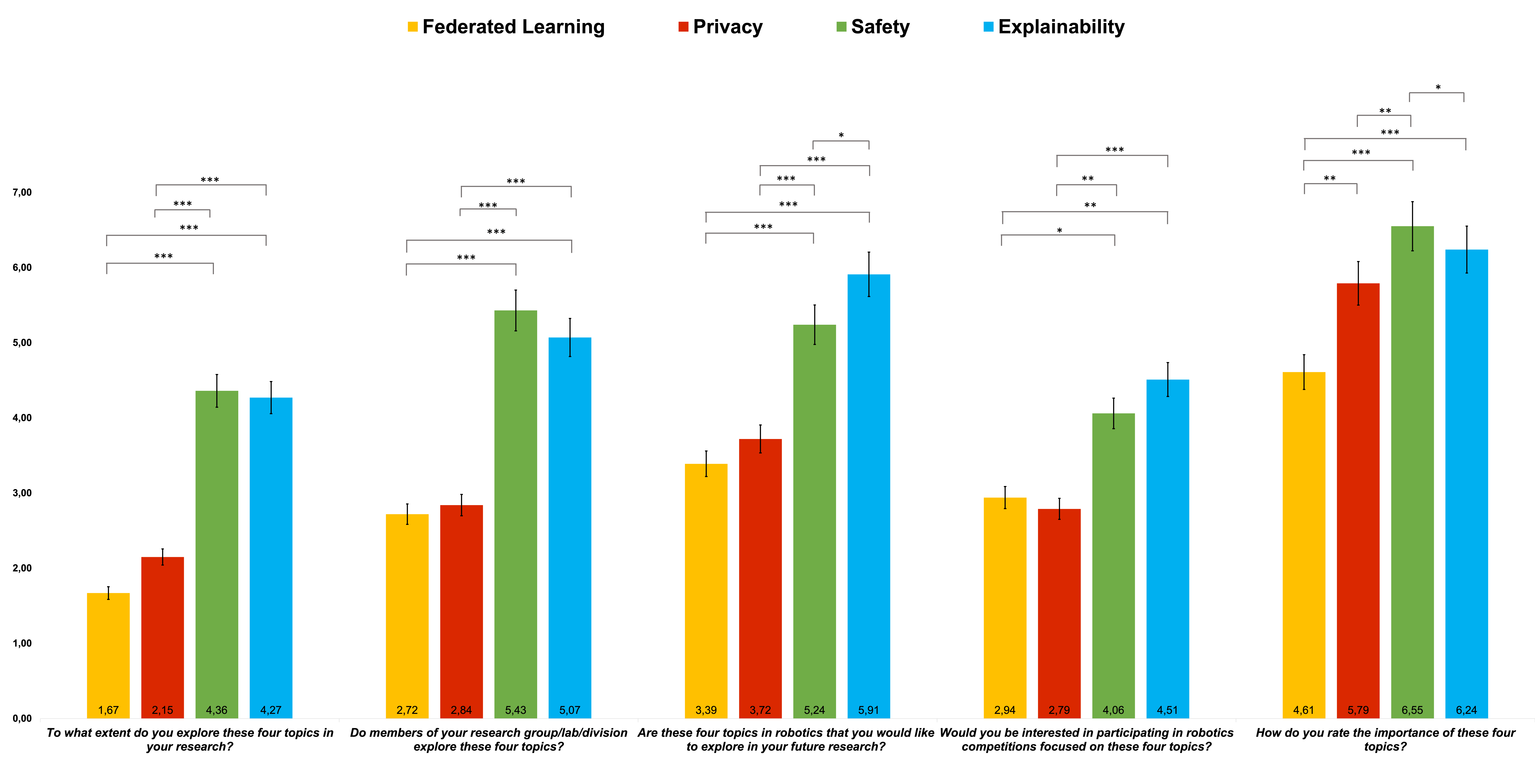}
  \caption{Results of collected responses from the questions using Likert scale (\textit{\textbf{Q1--Q5}}). Pairwise significance is displayed on the top. $*: p < 0.05$; $**: p < 0.005$; $***: p < 0.001$.}
  \label{fig:likert}
\end{figure}

To conduct the statistical analysis on the data obtained from the quantified scales, we performed the necessary normal distribution tests. The results confirmed that the data from \textit{\textbf{Q1--Q5}} all exhibited a normal distribution. Therefore, to examine the significant effects, the one-way repeated-measures analysis of variance (rmANOVA) \cite{zhang2021affective} was implemented for the answers from these questions. If any statistically significant effects (p < 0.05) disclosed, the Bonferroni-corrected post hoc tests were then performed to determine which pairs of the four topics were significantly different. All of the analysis was performed through IBM SPSS Statistics. Figure ~\ref{fig:likert} displays the descriptive results (mean, SD) and significance representation with 95\% confidence intervals (CI).

For \textit{\textbf{Q1}}, the results of the one-way rmANOVA with a Greenhouse-Geisser correction determined a significant main effect of the extent our participants' exploration among the four topics ($F$(2.572, 82.296) = 27.722, $p$ < 0.001). Moreover, Post hoc analysis with a Bonferroni adjustment revealed that statistical significance was eve found in some pairwise comparisons (Figure ~\ref{fig:likert}). For example, it indicates the extent that participants devoted towards \textit{federated learning} is significantly much less than that on \textit{safety}. Likewise, the rmAVONA a Sphericity Assumed correction showed that the significance is also found in the likelihood of the ambient colleagues explore the four topics  (\textit{\textbf{Q2}})($F$(3, 66) = 20.989, $p$ < 0.001). It is noteworthy some pairs of the topics again possessed significance through the Bonferroni post hoc tests. For instance, our participants' colleagues show a higher likelihood of exploring explainability compared to privacy, and this difference is statistically significant.  Noteworthily, our observation indicates that the intent for future exploration of the four topics was quantified with relatively high values (\textit{\textbf{Q3}}). Also, the rmANOVA with a Sphericity Assumed correction showed the significance was identified among the topics ($F$(3, 96) = 23.072, $p$ < 0.001), with more pairwise comparisons being significant (safety-explainability here) after the post hoc test. However, we found that the future interest (\textit{\textbf{Q4}}) in participating in robotics competitions received the lowest scores, showing a distinct decline across each topic compared to other questions. \textit{Federated learning} and \textit{privacy} gained particularly negative assessments, while \textit{safety} and \textit{explainability} elicited slightly more positive responses, although the average values merely approached neutrality (Scale: 4). Still, the rmANOVA with a Greenhouse-Geisser correction determined the existence of significance on the topics ($F$(2.551, 81.640) = 13.820, $p$ < 0.001), with some pairwise significance via the post hoc measurement. Through \textit{\textbf{Q5}}, we obtained an overview into how mainstream robotics researchers/practitioners perceive the four tested topics. \textit{Safety} received the highest score, while \textit{federated learning }was rated as the least important. Similarly, significance was identified from the rmANOVA with a Sphericity Assumed correction ($F$(3, 96) = 27.235, $p$ < 0.001), while post hoc pairwise comparison showed \textit{safety} was significantly deemed to be of more importance than explainability.

\subsection{Qualitative Results}
In this section, we present the qualitative responses from our survey study involving 34 participants. These insights, gleaned from the narratives and perspectives of our diverse participant pool, provide the opportunity for better understanding the perspectives regarding the four human-centric topics which are of interest to the mainstream robotics/HRI community. Beyond statistical analysis, the qualitative findings displayed here offer an additional deep and textured understanding that informs the public about the future exploration of essential design considerations in relevant competitions.

\textbf{\textit{Safety is widely acknowledged for its paramount significance}}\\
It emerged from our collected results that \textit{safety} is the most prioritized topics manifested either in the current research landscape or future research considerations in terms of the robotics community. One participant commented: \textit{"For embodied robots, the safety issue needs to be considered from the outset."} Most of the robotics researchers/practitioners in our study expressed the desire of prioritizing the human-robot safety so as to avoid any unwanted consequences, for instance, a collision which might jeopardize the human health. Similarly, another participant exemplified the navigation problem: \textit{"Safety applies to navigation as well. Instances such as abrupt stops or prolonged delays can pose challenges, especially in environments where people anticipate human-like behaviors."} One participant indicated the necessity of \textit{safety} in competitions, that it might be resolved by problem changing by time: \textit{"In competitions, it is crucial that problems evolve or broaden annually to prevent a competition from becoming a benchmark-focused pursuit that diverges from addressing real-world problems."}

\textbf{\textit{Explainability may be overlooked, but we are aware of its significance and difficulties}}\\
As displayed in the quantitative results, \textit{explainability} has gained less attention compared to \textit{safety} pertaining to the current research status, however, it was more recognized with respect to future explorations (Fig. ~\ref{fig:likert}). One participant pointed out: \textit{"Explainability, from my perspective, is an underappreciated aspect, and any initiative to highlight its importance is commendable. This becomes especially crucial in the current trajectory towards a black-box AI environment."} We found that the awareness and recognition of this topic is being noticed with the emergence of explainable AI \cite{gunning2019xai}. Another two comments we received (\textit{"I would anticipate that more researchers will delve into this explainability in the future."} and \textit{"The exaplainability will be intriguing in the exploration of comprehending agency in HRI."}) also imply that the acknowledge of this topic rooted with human centrism. In the meantime, our findings revealed that there are challenges and obstacles. One participant reported: \textit{"Achieving excellence in explainability appears to be challenging, given its reliance on human comprehension of the explanations."}, which reflects that making explanations accessible needs to be considered and prioritized. In addition, one participant posed the concerns of engaging parallel topics to obtain desirable explainability: \textit{"Besides, transparency should also be incorporated at the same time."}

\begin{figure}[ht!]
  \centering
  \includegraphics[width=.8\columnwidth]{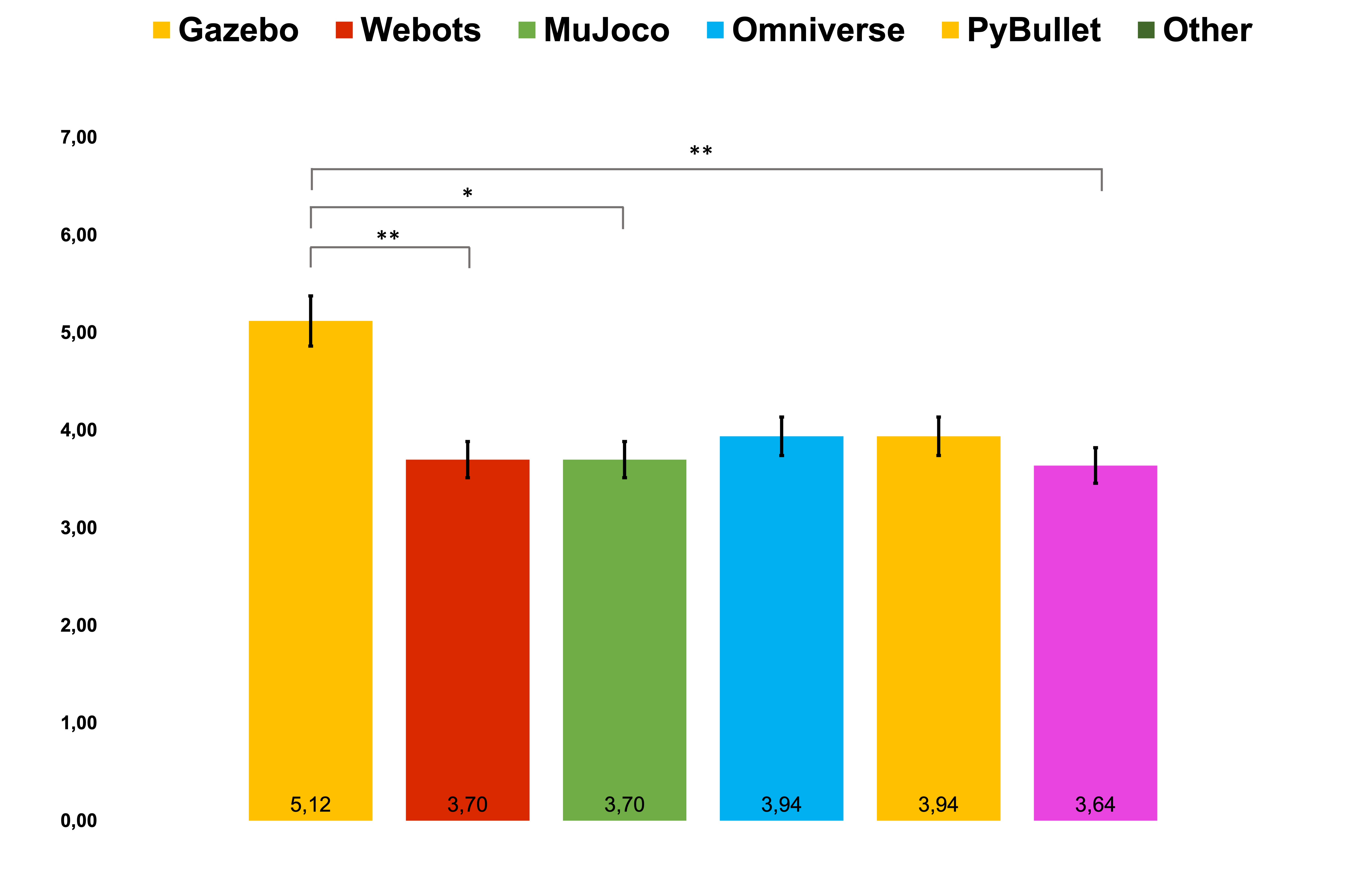}
  \caption{The comparison of participant's preferences on robotic simulators.}
  \label{fig:simulator}
\end{figure}

\textbf{\textit{Privacy is important, but yet hard to realize it}}\\
Similarity has been identified between qualitative reasoning and quantitative analysis regarding privacy, that it receives considerably less recognition within both the present research community and prospective explorations in contrast of \textit{safety} and explainability. Many participants reported that \textit{privacy} should be valued since it relates with ethical issues, for example, one commented: \textit{"I think privacy holds considerable importance for robots due to that there would be multitude of sensors equipped with. It is imperative to establish explicit guidelines for data flow and ownership."} Nonetheless, the most common issue appeared was the tremendous difficulties including lack of prior knowledge which hinder successful exploitation of privacy in robotics and prospective competitions. \textit{"Privacy consistently ranks as one of the foremost concerns; however, addressing it in a robotics competition necessitates a thoughtful setup to imbue it with meaningful contexts"}, stated by one participant, similar to another: \textit{"I have no ability to envision its viability and feasibility within a competition setting."} Moreover, one specifically concerned about the limited generalizability which would be formidable: \textit{"This topic is more broadly associated with AI as a whole, extending beyond the scope of just robotics."}

\textbf{\textit{Federated learning is not popular, and should not be separated from privacy}}\\
Through our results, we disclosed that so far, \textit{federated learning} has the least attention among the robotics community (also refer to Figure ~\ref{fig:likert}). Most of the qualitative feedback we received showed that besides the unpopularity, the infeasibility of federated learning in a robotics competition is the moat influential factor (\textit{"At first glance, implementing authentic federated learning in a competitive environment appears very challenging."; \textit{"From my perspective, I'm skeptical about its practicality within the context of a robotics competition."}}). Furthermore, we recognized the inherent association of federated learning with privacy, emphasizing the importance of maintaining this connection rather than detachment, in terms of the robotics community. for example, one participant \textit{"believes that federated learning cannot and should not be disentangled from privacy considerations".} The finding coincides with what we mentioned in Section ~\ref{intro}: as an advanced machine learning mechanism, federated learning is privacy-preserving in order to achieve no inference in sensitive information.

\begin{figure}[ht!]
  \centering
  \includegraphics[width=.7\columnwidth]{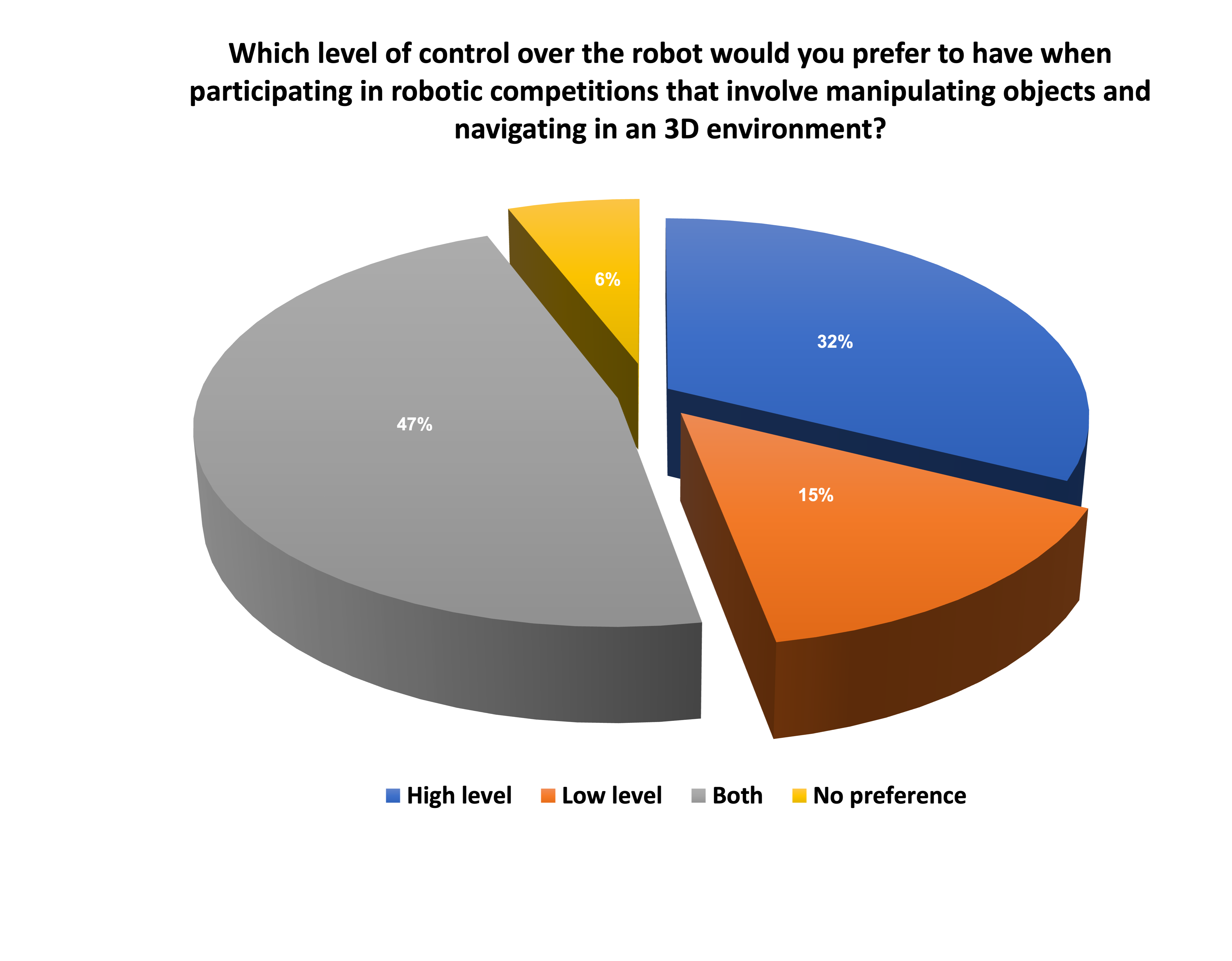}
  \caption{Participants' preferences on high/low level control in robotics competitions.}
  \label{fig:control}
\end{figure}

\subsection{Robotic Control and Simulators}
In addition, we analyzed both the preference over the type of robotic simulator employed in robotic competitions and the level of control of the robotic agent in simulation. We considered five different robotic simulators, widely used by the community: Gazebo, an open-source simulator based on the ROS platform; Webots \cite{michel2004cyberbotics}, an open-source 3D simulator; MuJoco \cite{todorov2012mujoco}, an open-source physics simulator; Omniverse, a proprietary robotic simulator; and PyBullet, an open-source and python-based physics engine. The results are presented in Figure ~\ref{fig:simulator}. The results show a clear preference for the Gazebo platform, a more established platform, over more recent platforms such as MuJoco. This insight is expected given the experience level of the participants in the study.

In addition, the participants were also asked about their preferred level of control of the robot agent in the simulation environment. The participants could select low-level control, in which the roboticist can have control over the low-level joint and motor sensors of the robot, high-level control, in which roboticist can only control high-level actions of the robotic platform (e.g., move to a specified position, pick up an object), or both. Moreover, the participants could also select they had no preference over the level of control. The results, shown in Figure ~\ref{fig:control}, show a preference for high-level or both level of control, in comparison to low-level control.

\section{Discussion}
\label{dis}

\subsection{Insights}
We conducted a needfinding study to initially identify the current research landscape of the the robotics community towards four human-centric topics -- \textit{safety}, \textit{privacy}, \textit{explainability}, and \textit{federated learning}. Moreover, we sought to probe the intention of robotics researchers/practitioners in participating these four themed competitions. A questionnaire was then formulated and answered by 34 recruited participants from three renowned robotics consortia across Europe. Close to 60\% of the participants possess over five years of experience in robotics and related fields, which renders our sample group representative and compelling. The findings suggest that the mainstream robotics community holds distinct viewpoints on the four topics, along with varying attitudes toward participating in competitions related to each topic.

What do we learn from the study? In both current research status and future research agenda, \textit{safety} and \textit{explainability} elicit significantly greater concern compared to \textit{privacy} and federated learning. On one hand, \textit{safety} emerges as a more prevalent focus in current robotics research exploration. On the other hand, \textit{explainability} is regarded as a more favored topic in future research directions. This trend is attributed to the constant consideration of safety as an imperative issue in the design of robotic/HRI systems, as reported by participants. Additionally, our results highlighted an increasing focus on anticipating robotic behaviors to be explainable, marking it as a central concentration in the past years. In comparison, \textit{privacy} and \textit{federated learning} are much less noticed at present and less possible being considered in the future research. Particularly, \textit{federated learning} has the least exposure among the four topics. This could be summarized to two main factors: (1) Persistent challenges in delving into privacy-related areas due to the sensitivity of data; (2) The mainstream community has not yet universally embraced federated learning.

An evident quantification decrease was observed from our results regarding the interest in participating the four themed robotics competitions. In a word, this reveals a lack of strong enthusiasm within the robotics community for participating in future competitions focused on these four human-centric topics. This may because of the deficiency of the previous experience in similar competitions (Figure~\ref{fig:prior}.$b$). In addition, the uncertainty and unforeseen risk of participating a robotic competition can be another obstacle, since substantial dedication of effort is frequently required in such a context where real robots might be needed. For example, testing practical robots may demand precision and a considerable amount of endeavour, making it a time-consuming process. An intriguing fact is that despite a low willingness for competition participation, \textit{explainability} was ranked as the most desired topic, whereas \textit{privacy} was treated as the least favored.We believe this can be ascribed to: (1) The growing interest in researching explainable AI; (2) Lack of robotics competitions focusing on privacy, as the authors conducted a comprehensive search and found no such competitions held in prestigious robotics venues so far. As these four topics are all closely relevant to AI, we expect that the target group can be shifted towards the machine learning/AI community.

\subsection{Limitations}
We acknowledge the presence of some certain limitations in our study Firstly, the survey was mostly implemented online, while physical interaction with more constructed interviews might provide more in-depth insights and feedback. Secondly, the questions we designed for the users may not incorporate a sufficiently comprehensive scope to thoroughly examine the attitudes of the robotics community, potentially impacting the depth of the results. Last but not least, the participants in our study were exclusively based in Europe, which may deviate from the inclusive principles commonly employed in defining sample groups in user study design.

\section{Conclusion}
\label{con}
In this paper, we proposed a needfinding study to investigate the prospective enthusiasm of the mainstream robotics community towards the participation in future competitions themed with four human-centric topics. We found that a low level of willingness was observed according to the results, implicating the possible change of the target community. We envision the needfinding study presented in this paper to be as an informative reference for robotic researchers/practitioners to learn the exploratory space and get inspired prior to participating or holding the pertinent robotics competitions. We hope our investigation can offer advisable guidance for the robotics/HRI community in advancing the new frontiers of the related research, especially future competitions considerations in \textit{safety}, \textit{privacy}, \textit{explainability}, and \textit{robust federated learning}.


\section*{Acknowledgment}
This work was funded by the HORIZON-CL4-2021-HUMAN-01 ELSA project and the Swedish Foundation for Strategic Research (SSF) grant FUS21-0067.
%
%
%
\bibliographystyle{splncs04}
\bibliography{mybibliography}

\end{document}